\pdfoutput=1
\documentclass[12pt,a4paper]{article}

\usepackage{ifthen} 
\newboolean{pdflatex}
\setboolean{pdflatex}{true} 

\newboolean{articletitles}
\setboolean{articletitles}{true} 

\newboolean{uprightparticles}
\setboolean{uprightparticles}{false} 

\def\paperauthors{Neville Harnew$^1$, for the LHCb collaboration} 
\def\paperasciititle{LHCb Upgrade II: status and physics prospects} 
\def\papertitle{LHCb Upgrade II: status and physics prospects} 
\def\paperkeywords{{High Energy Physics}, {LHCb}} 
\def\papercopyright{\the\year\ CERN for the benefit of the LHCb collaboration} 
\def\paperlicence{CC BY 4.0 licence}
\def\paperlicenceurl{https://creativecommons.org/licenses/by/4.0/}

\newif\ifEnableSectionTOCLinks
\EnableSectionTOCLinksfalse 

\usepackage[top=1in, bottom=1.25in, left=1in, right=1in]{geometry}

\usepackage{microtype}
\usepackage{lineno}  
\usepackage{xspace} 
\usepackage{caption} 

\usepackage{graphicx}  
\usepackage{color}
\usepackage{colortbl}
\graphicspath{{./figs/}} 

\usepackage{amsmath} 
\usepackage{amssymb}
\usepackage{amsfonts}
\usepackage{upgreek} 

\usepackage[normalem]{ulem} 

\newcommand*\patchAmsMathEnvironmentForLineno[1]{%
\expandafter\let\csname old#1\expandafter\endcsname\csname #1\endcsname
\expandafter\let\csname oldend#1\expandafter\endcsname\csname
end#1\endcsname
 \renewenvironment{#1}%
   {\linenomath\csname old#1\endcsname}%
   {\csname oldend#1\endcsname\endlinenomath}%
}
\newcommand*\patchBothAmsMathEnvironmentsForLineno[1]{%
  \patchAmsMathEnvironmentForLineno{#1}%
  \patchAmsMathEnvironmentForLineno{#1*}%
}
\AtBeginDocument{%
\patchBothAmsMathEnvironmentsForLineno{equation}%
\patchBothAmsMathEnvironmentsForLineno{align}%
\patchBothAmsMathEnvironmentsForLineno{flalign}%
\patchBothAmsMathEnvironmentsForLineno{alignat}%
\patchBothAmsMathEnvironmentsForLineno{gather}%
\patchBothAmsMathEnvironmentsForLineno{multline}%
\patchBothAmsMathEnvironmentsForLineno{eqnarray}%
}

\usepackage[pdftex,
            pdfauthor={\paperauthors},
            pdftitle={\paperasciititle},
            pdfkeywords={\paperkeywords}]{hyperref}
\usepackage{hyperxmp}
\hypersetup{
    pdfcopyright={Copyright (C) \papercopyright},
    pdflicenseurl={\paperlicenceurl}
}

\usepackage[colorinlistoftodos,textsize=scriptsize]{todonotes}

\usepackage[bottom,flushmargin,hang,multiple]{footmisc}

\usepackage[all]{hypcap} 

\usepackage{xspace} 
\usepackage{upgreek}

\def\MagUp {\mbox{\em Mag\kern -0.05em Up}\xspace}

\ifdefined\ifuprightparticles
\else
\newboolean{uprightparticles}
\setboolean{uprightparticles}{false} 
\fi

\ifthenelse{\boolean{uprightparticles}}%
{

 \def\PDelta      {\ensuremath{\Delta}\xspace}                 
 \def\PXi         {\ensuremath{\Xi}\xspace}                 
 \def\PLambda     {\ensuremath{\Lambda}\xspace}                 
 \def\PSigma      {\ensuremath{\Sigma}\xspace}                 
 \def\POmega      {\ensuremath{\Omega}\xspace}                 
 \def\PUpsilon    {\ensuremath{\Upsilon}\xspace}
 \let\oldPi\Pi
 \def\PPi         {\ensuremath{\oldPi}\xspace}

 \def\PB      {\ensuremath{\mathrm{B}}\xspace}                 
 \def\PD      {\ensuremath{\mathrm{D}}\xspace}                 

 \def\PK      {\ensuremath{\mathrm{K}}\xspace}                 
 \def\Ps      {\ensuremath{\mathrm{s}}\xspace}

 \def\thebaroffset{0.0em}
}
{

 \mathchardef\PDelta="7101
 \mathchardef\PXi="7104
 \mathchardef\PLambda="7103
 \mathchardef\PSigma="7106
 \mathchardef\POmega="710A
 \mathchardef\PUpsilon="7107
 \mathchardef\PPi="7105
 \def\PB      {\ensuremath{B}\xspace}                 
 \def\PD      {\ensuremath{D}\xspace}                 

 \def\PK      {\ensuremath{K}\xspace}                 
 \def\Ps      {\ensuremath{s}\xspace}

 \def\thebaroffset{0.18em}
}
\newcommand{\offsetoverline}[2][\thebaroffset]{\kern #1\overline{\kern -#1 #2}}%

\makeatletter
\ifcase \@ptsize \relax
  \newcommand{\miniscule}{\@setfontsize\miniscule{4}{5}}
\or
  \newcommand{\miniscule}{\@setfontsize\miniscule{5}{6}}
\or
  \newcommand{\miniscule}{\@setfontsize\miniscule{5}{6}}
\fi
\makeatother

\DeclareRobustCommand{\optbar}[1]{\shortstack{{\miniscule (\rule[.5ex]{1.25em}{.18mm})}
  \\ [-.7ex] $#1$}}

\def\squark    {{\ensuremath{\Ps}}\xspace}

\def\KorKbar {\kern \thebaroffset\optbar{\kern -\thebaroffset \PK}{}\xspace}

\def\D       {{\ensuremath{\PD}}\xspace}

\def\DorDbar {\kern \thebaroffset\optbar{\kern -\thebaroffset \PD}\xspace}

\def\Dp      {{\ensuremath{\D^+}}\xspace}
\def\Dm      {{\ensuremath{\D^-}}\xspace}

\def\DpDm    {\ensuremath{\Dp {\kern -0.16em \Dm}}\xspace}

\def\B       {{\ensuremath{\PB}}\xspace}

\def\BorBbar {\kern \thebaroffset\optbar{\kern -\thebaroffset \PB}\xspace}

\def\Bd      {{\ensuremath{\B^0}}\xspace}

\def\BdorBdbar {\kern \thebaroffset\optbar{\kern -\thebaroffset \Bd}\xspace}

\def\Bs      {{\ensuremath{\B^0_\squark}}\xspace}

\def\BsorBsbar {\kern \thebaroffset\optbar{\kern -\thebaroffset \Bs}\xspace}

\def\Y#1S{\ensuremath{\PUpsilon{(#1S)}}\xspace}

\def\LorLbar     {\kern \thebaroffset\optbar{\kern -\thebaroffset \PLambda}\xspace}

\def\AT#1     {\ensuremath{A_{\mathrm{T}}^{#1}}\xspace}           

\def\C#1      {\ensuremath{\mathcal{C}_{#1}}\xspace}                       
\def\Cp#1     {\ensuremath{\mathcal{C}_{#1}^{'}}\xspace}                    
\def\Ceff#1   {\ensuremath{\mathcal{C}_{#1}^{\mathrm{(eff)}}}\xspace}        
\def\Cpeff#1  {\ensuremath{\mathcal{C}_{#1}^{'\mathrm{(eff)}}}\xspace}       
\def\Ope#1    {\ensuremath{\mathcal{O}_{#1}}\xspace}                       
\def\Opep#1   {\ensuremath{\mathcal{O}_{#1}^{'}}\xspace}                    

\newcommand{\aunit}[1]{\ensuremath{\text{\,#1}}}       

\newcommand{\tev}{\aunit{Te\kern -0.1em V}\xspace}
\newcommand{\gev}{\aunit{Ge\kern -0.1em V}\xspace}
\newcommand{\mev}{\aunit{Me\kern -0.1em V}\xspace}
\newcommand{\kev}{\aunit{ke\kern -0.1em V}\xspace}
\newcommand{\ev}{\aunit{e\kern -0.1em V}\xspace}
 
\newcommand{\mevc}{\ensuremath{\aunit{Me\kern -0.1em V\!/}c}\xspace}
\newcommand{\gevc}{\ensuremath{\aunit{Ge\kern -0.1em V\!/}c}\xspace}
\newcommand{\mevcc}{\ensuremath{\aunit{Me\kern -0.1em V\!/}c^2}\xspace}
\newcommand{\gevcc}{\ensuremath{\aunit{Ge\kern -0.1em V\!/}c^2}\xspace}

\def\gsim{{~\raise.15em\hbox{$>$}\kern-.85em
          \lower.35em\hbox{$\sim$}~}\xspace}
\def\lsim{{~\raise.15em\hbox{$<$}\kern-.85em
          \lower.35em\hbox{$\sim$}~}\xspace}

\def\tell1  {TELL1\xspace}
\def\ukl1   {UKL1\xspace}

\newcommand{\lhcborcid}[1]{\href{https://orcid.org/#1}{\hspace*{0.1em}\raisebox{-0.45ex}{\includegraphics[width=1em]{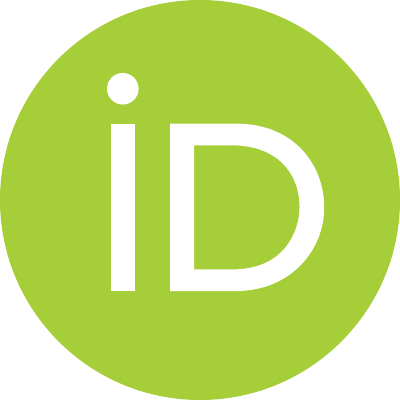}}}}

\hypersetup{
  colorlinks   = true, 
  urlcolor     = blue, 
  linkcolor    = blue, 
  citecolor    = red   
}

\ifEnableSectionTOCLinks
    \usepackage[explicit]{titlesec} 
    
    \let\oldcontentsline\contentsline
    \renewcommand

    \titleformat{\section}{\normalfont\Large\bf}{\hyperlink{tocsection.\thesection}{{\thesection} \parbox[t]{\dimexpr\textwidth-1pc}{#1}}}{1pc}{}

    \titleformat{\subsection}{\normalfont\bf}{\hyperlink{tocsubsection.\thesubsection}{{\thesubsection} \parbox[t]{\dimexpr\textwidth-1pc}{#1}}}{1pc}{}

    \titleformat{name=\section,numberless}[display]{}{}{0pt}{\normalfont\Huge\bfseries #1}
\fi

\usepackage{cite} 
\usepackage{LHCb/mciteplus}

\usepackage{longtable} 

\begin{document}

\renewcommand{\thefootnote}{\fnsymbol{footnote}}
\setcounter{footnote}{1}


\begin{titlepage}

\vspace*{-1.5cm}

\noindent
\begin{tabular*}{\linewidth}{lc@{\extracolsep{\fill}}r@{\extracolsep{0pt}}}
\ifthenelse{\boolean{pdflatex}}
{\vspace*{-1.2cm}\mbox{\!\!\!\includegraphics[width=.14\textwidth]{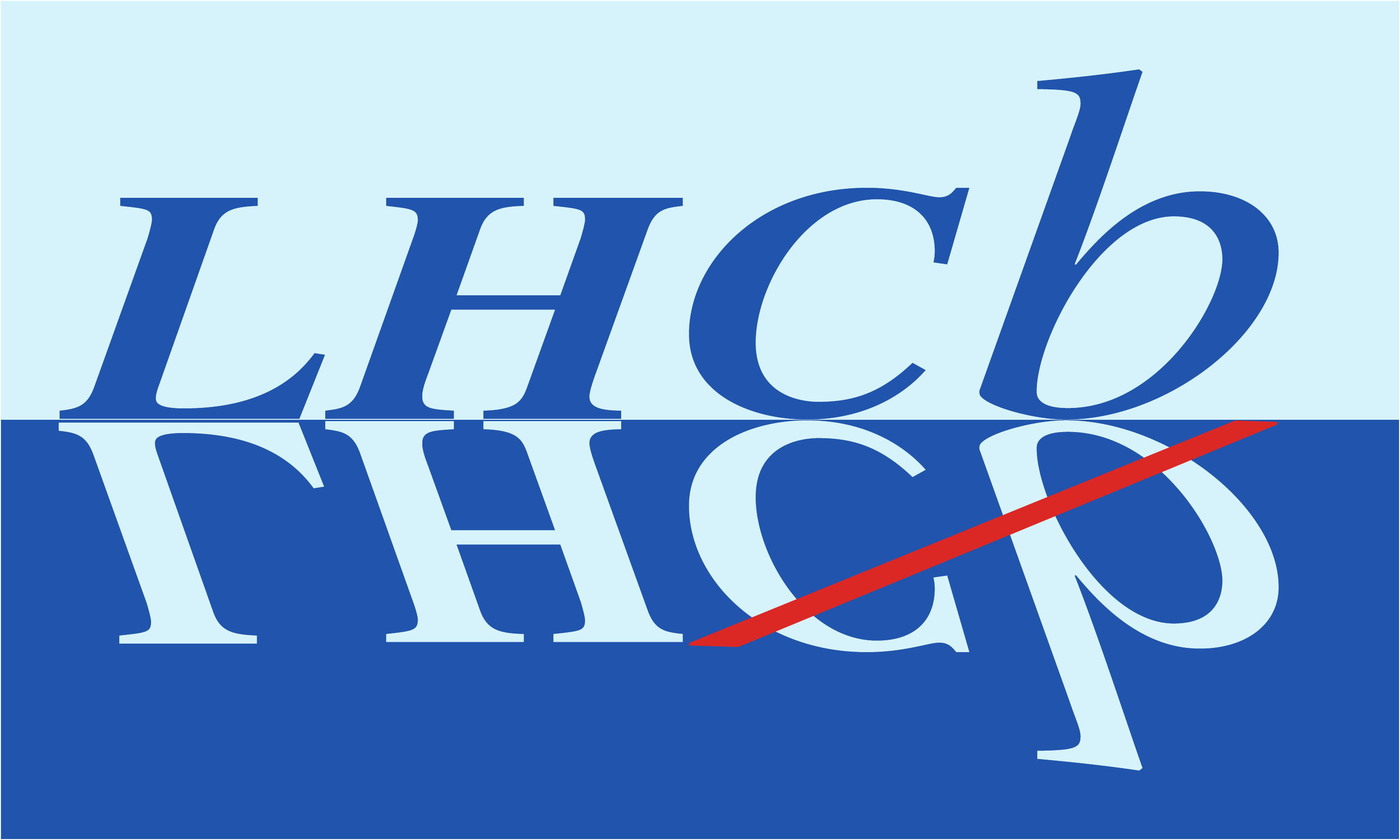}} & &}%
{\vspace*{-1.2cm}\mbox{\!\!\!\includegraphics[width=.12\textwidth]{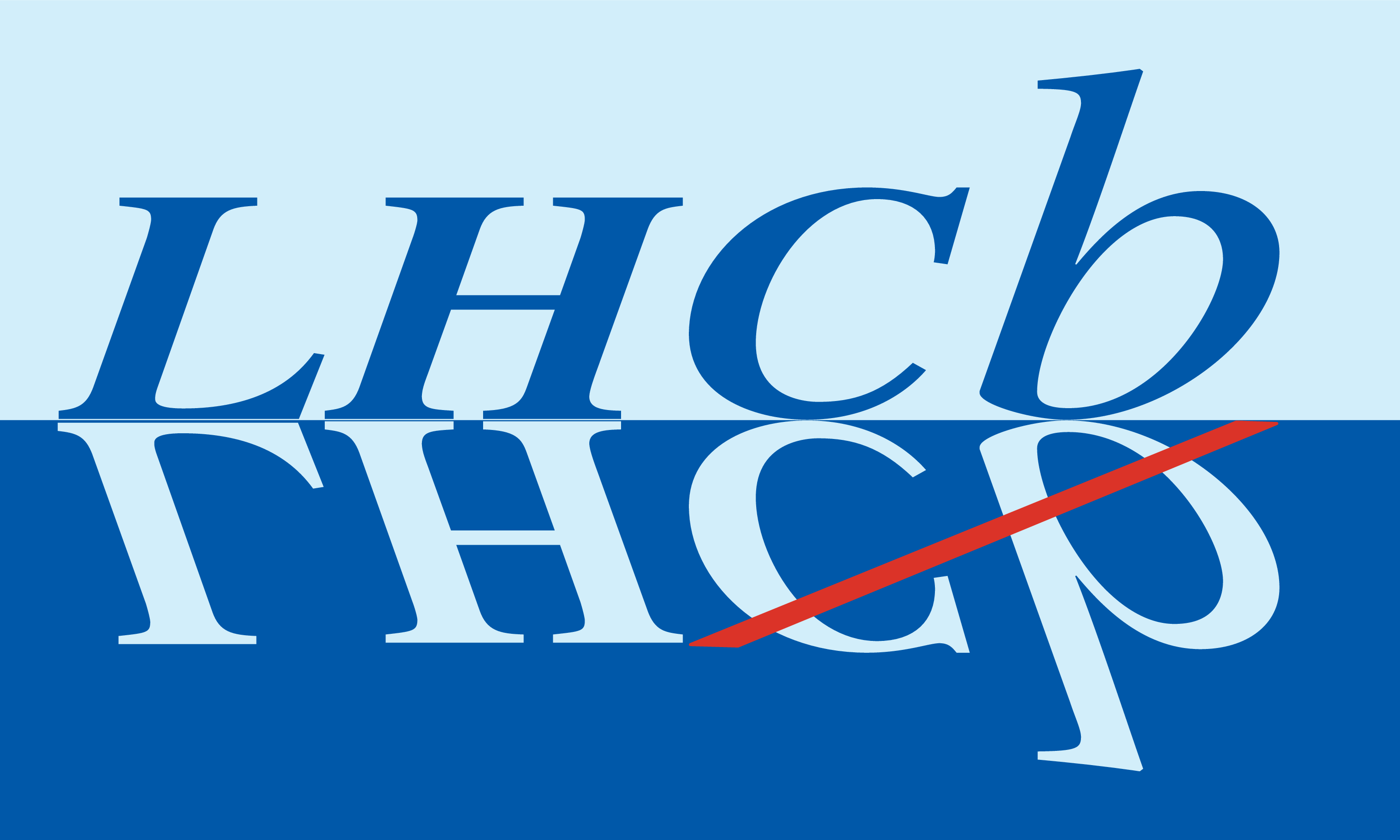}} & &}
 \\
 & & LHCb-PROC-2026-002 \\  
 & & \today \\ 
 & & \\
\hline
\end{tabular*}

\vspace*{4.0cm}

{\normalfont\bfseries\boldmath\huge
\begin{center}
  \papertitle
\end{center}
}

\vspace*{2.0cm}

\begin{center}
\paperauthors.
\bigskip\\
{\normalfont\itshape\footnotesize
$ ^1$Denys Wilkinson Building, Department of Physics, Keble Road, Oxford OX1 3RH, United Kingdom\\
}
\end{center}

\vspace{\fill}

\begin{abstract}
  \noindent This paper summarises the status of the LHCb Upgrade II project for  operation at the High Luminosity LHC.
The physics motivation to reach an integrated luminosity of \(300\text{ fb}^{-1}\), the sub-detector designs, the project schedule, the detector scoping, and planning towards data-taking will be reviewed.  

\end{abstract}

\vspace*{7.0cm}
\vspace{\fill}
{\footnotesize
\centerline{\copyright~\papercopyright. \href{\paperlicenceurl}{\paperlicence}.}}
\vspace*{2mm}

\begin{center}
   \footnotemark[0]
   \footnotetext[0]{Conference proceedings prepared for the 21st International Conference on {\ensuremath{B}}--Physics at Frontier Machines (Beauty2026), Maastricht, Netherlands,
1--5 June, 2026. \\
   Email:  Neville Harnew, 
   \href{mailto:Neville.Harnew@physics.ox.ac.uk}{Neville.Harnew@physics.ox.ac.uk}.
 }
\end{center}

\vspace{\fill}

\end{titlepage}

\pagestyle{empty}  


\newpage
\setcounter{page}{2}
\mbox{~}


\renewcommand{\thefootnote}{\arabic{footnote}}
\setcounter{footnote}{0}


\cleardoublepage


\pagestyle{plain} 
\setcounter{page}{1}
\pagenumbering{arabic}


\section{Introduction}

The LHCb experiment is a single-arm forward spectrometer operating at the Large Hadron Collider (LHC) which has   successfully collected data during Runs 1 and 2 of the LHC\,\cite{LHCb:2014mvo}. The detector underwent a major overhaul (Upgrade I) prior to the start of LHC Run 3\,\cite{LHCb:2024upgrade} which has now ended. Run 3  operation was at a running luminosity of approximately \mbox{$\sim 2\,\times$\,10$^{33}$\,cm$^{-2}$s$^{-1}$}  featuring greatly improved tracker and particle identification (PID) systems from Runs 1 and 2, alongside the first high-throughput GPU trigger in a high-energy physics  experiment. This trigger performed a near-full detector reconstruction with the added benefit of high efficiency, in particular for hadronic decay modes. At the \mbox{Upgrade I}  luminosity,  a production rate of approximately \(\mathcal{O}(10^6)\) \(b\bar{b}\)  quark pairs per second was achieved, with the corresponding \(c\bar{c}\) production cross-section yielding a  rate roughly twenty times larger.   The corresponding accumulated integrated luminosity since the start of \mbox{Run 1} is now 37\,pb$^{-1}$, approximately 50\% above original Upgrade I expectations.
Looking forward, preparations are now underway for the experiment to transition to \mbox{Upgrade II}, operating at a nominal instantaneous luminosity of \(1\times 10^{34} \text{ cm}^{-2}\text{s}^{-1}\) at the High Luminosity LHC (HL-LHC). 

This paper summarises the LHCb Upgrade II status and is organized as follows.
Section\,2 outlines the physics motivation for Upgrade II to reach a target integrated luminosity of \(300\text{ fb}^{-1}\). 
Section\,3 reviews the  status of sub-detector designs with a focus   on the challenges of the high pile-up environment which requires radiation-hard front-end readout incorporating timing. 
 Section\,4 outlines the project schedule, scoping and planning towards data-taking, whilst Section\,5 provides a short summary.

\section{Physics motivation for Upgrade II}

Even though the physics programme of LHCb has been a spectacular success, the need to uncover physics beyond the Standard Model (SM)  demands an even more ambitious experimental strategy.  Conclusive and statistically-significant deviations from SM predictions have been elusive, and it is a primary goal of the European Strategy for Particle Physics to fully exploit the LHC’s potential towards new-physics discoveries\,\cite{CERN:ESU2026}.  Hence, looking forward, the experiment will evolve towards Upgrade II\,\cite{LHCb:2776420, LHCb:2024scoping}
which will  offer a unique opportunity to conduct indirect searches for new heavy particles at vastly higher mass scales than those accessible via direct production. The statistical improvement resulting from an anticipated integrated luminosity of \(300\text{ fb}^{-1}\) after ten years of running will enable significantly higher precision in measurements of CP-violation in beauty and charm decays and heavily-suppressed loop-level processes.

\begin{figure}[h]
  \begin{center}
    \includegraphics[width=0.49\linewidth]{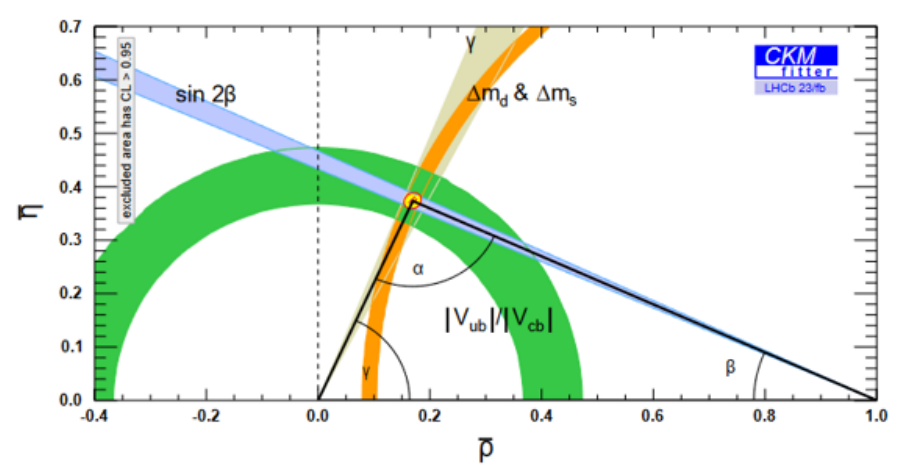} 
        \includegraphics[width=0.49\linewidth]{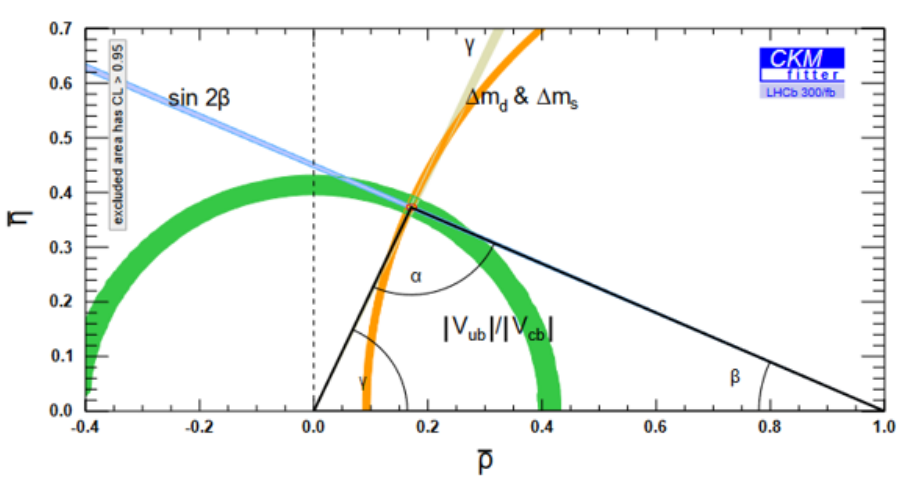} 
  \end{center}
  \caption{
Evolution of the Unitarity Triangle from (left) a 23\,fb$^{-1}$ LHCb dataset to (right) from Upgrade II after 300\,fb $^{-1}$\,\cite{LHCb:2018nre, Charles:2004jd}.
    }
  \label{fig:UTevolution}
\end{figure}

\begin{figure}[h]
  \begin{center}
    \includegraphics[width=0.85\linewidth]{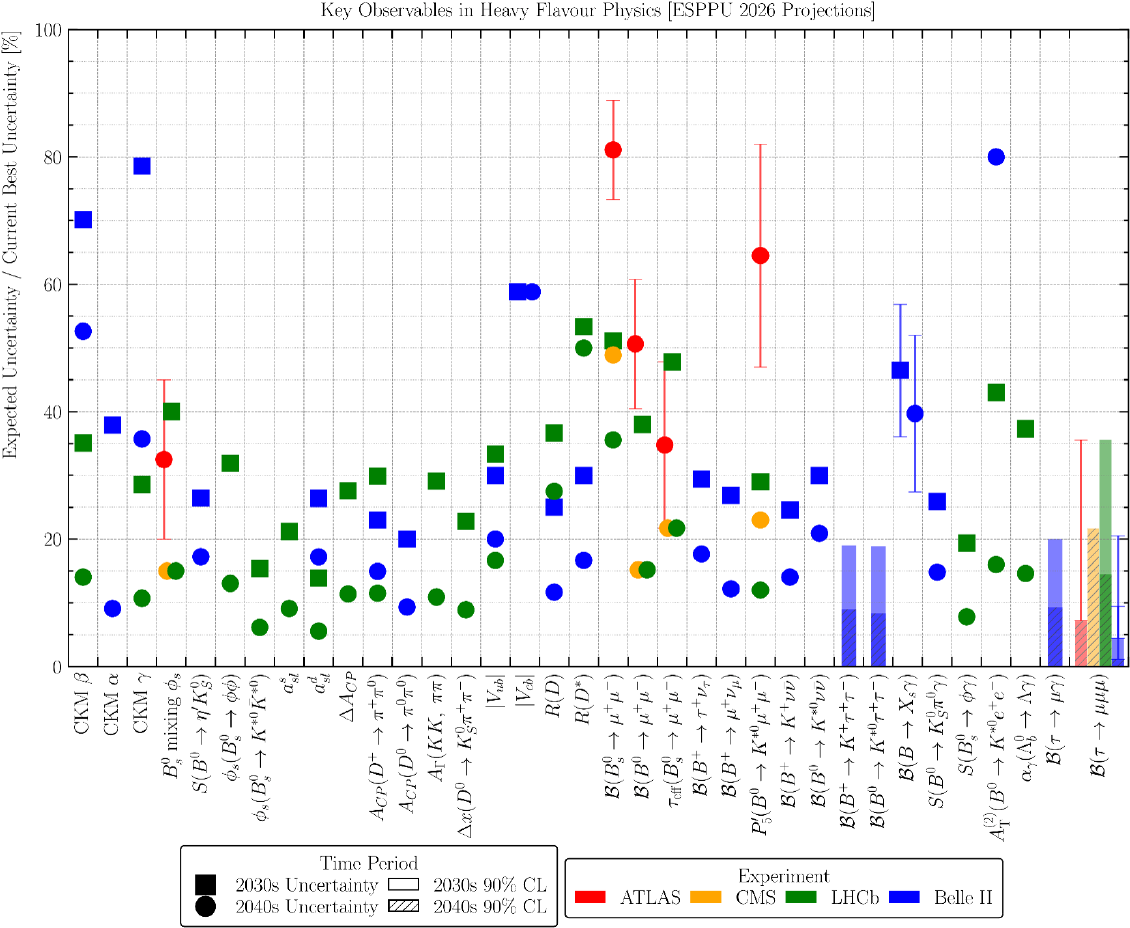} 
  \end{center}
  \caption{
The fractional percentage reduction of uncertainties of key measurements from 300\,fb\,$^{-1}$ of data from LHCb Upgrade II and other experiments,  with respect to the current best values\,\cite{ATLAS:2025xxv}.
    }
  \label{fig:performance}
\end{figure}

The expected evolution of Unitarity Triangle measurements from  an  LHCb Run\,3 dataset of 23\,fb$^{-1}$ compared to that from Upgrade II after 300\,fb $^{-1}$ is shown in Fig.\,\ref{fig:UTevolution}\,\cite{LHCb:2018nre, Charles:2004jd}.  Both plots correspond to  measurements from LHCb only, the latter also assuming improved theoretical input.
The improvements in precision of benchmark measurements after 10 years and 300\,fb$^{-1}$  of LHCb Upgrade II running 
are represented in Fig.\,\ref{fig:performance}\,\cite{LHCb:2018nre, ATLAS:2025xxv}. This plot shows the fractional percentage reduction of uncertainties of key measurements with respect to the current best values, also including those from other experiments.

\section{The LHCb Upgrade II detector}

\begin{figure}[h]
  \begin{center}
    \includegraphics[width=0.85\linewidth]{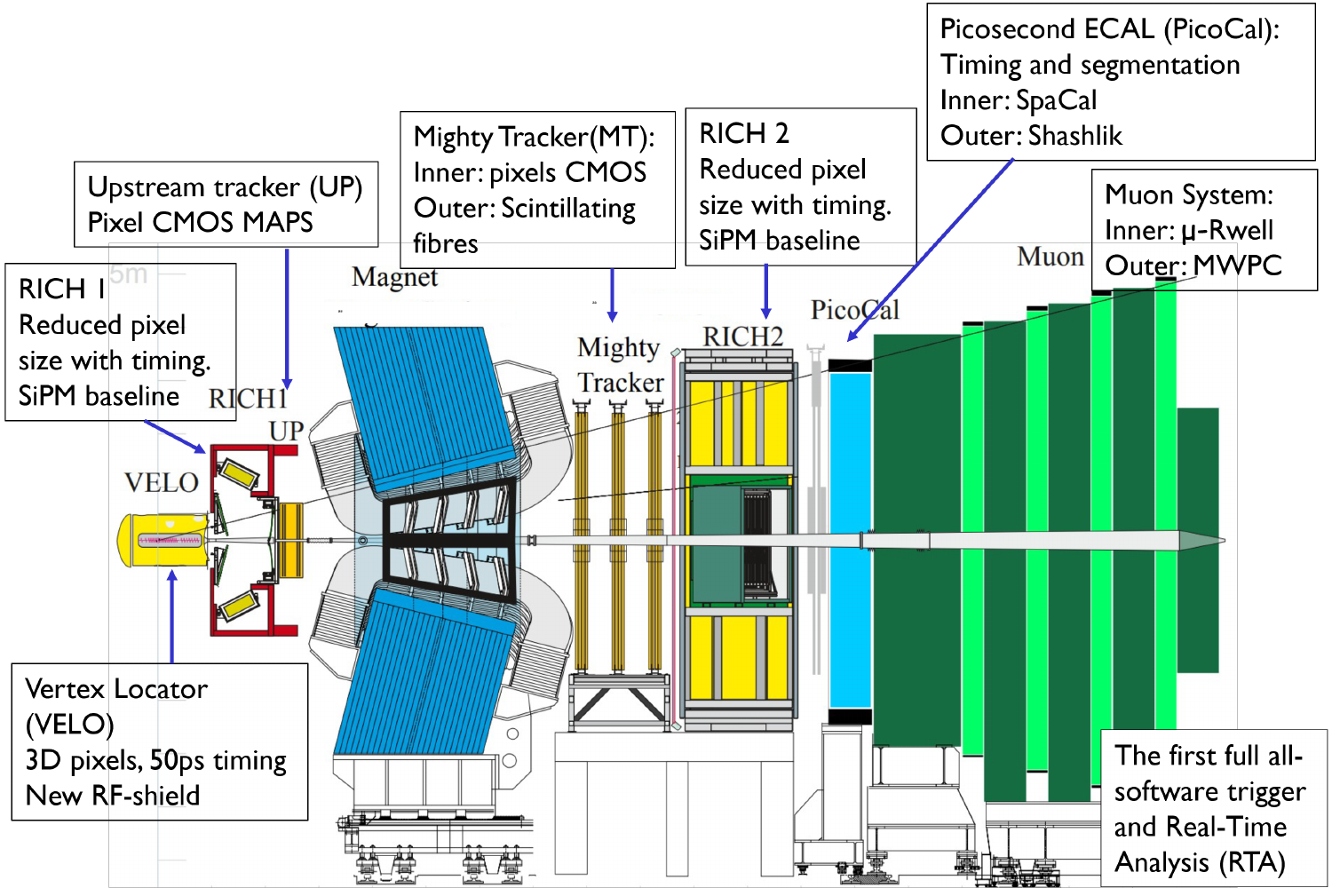} 
  \end{center}
  \caption{
The LHCb Upgrade II detector configuration. \cite{LHCb:2776420, LHCb:2024scoping}.
    }
  \label{fig:Upgrade-II-detector}
\end{figure}

The Upgrade II detector configuration \cite{LHCb:2776420, LHCb:2024scoping}
is shown in Fig.\,\ref{fig:Upgrade-II-detector}, which highlights the detector components to be described in this section.

To mitigate the substantial increase in the number of interactions per bunch crossing (pile-up) when the  previous luminosity of \(2\,\times\,10^{33}\,\text{cm}^{-2}\text{s}^{-1}\) is increased   to \(1\,\times\,10^{34}\,\text{cm}^{-2}\text{s}^{-1}\)  for Upgrade II, precise timing of the detector components will be key. The luminosity increase will produce approximately 2000 tracks on average  which results in an  escalation in pile-up, rising from approximately 5 interactions in Run 3, to 30 projected in \mbox{Run 5}. This requires improved detector granularity and the addition of an extra dimension, namely timing, which provides so-called four-dimensional (4D) reconstruction. 
Only by selecting tracks within a pre-defined time window of the beam-crossing, as an example \(\mathcal{O}(30)\)\,ps, will allow individual interactions to be isolated.

{\bf The tracking system} : 
The tracking system comprises three main components: the Vertex Locator, the Upstream Tracker and the Mighty Tracker, the latter incorporating both an inner and outer region. 
The system is designed to achieve a target momentum resolution of \(\Delta p/p \sim 0.4\text{--}0.5\%\) for tracks traversing the full length of the LHCb spectrometer.

The VErtex LOcator (VELO) :   To emulate the Run 3 performance for Upgrade II, the VELO is required to have an increased granularity with a pixel pitch of \(50 \times 50\,\mu\text{m}^2\) and a per-hit time resolution of \(\le 50\,\text{ps}\). The baseline design utilizes 3D silicon pixel sensors fabricated in \(28\,\text{nm}\) CMOS technology, featuring low capacitance to deliver a \(40\,\text{ps}\) timing resolution. The impact parameter resolution will be \(\sim 11\,\mu\text{m}\) for high transverse momentum tracks. The Technical Design Report (TDR) for the VELO is expected to be ready by the first quarter (Q1) of  2027.

Upstream Tracker (UP):   Serving as a crucial bridge between the VELO and the Mighty Tracker, the UP ensures high overall tracking performance and reduces the rate of ghost tracks. Whilst the Run 3 upstream tracker utilizes silicon strips, the  UP will have  pixels consisting of four planes of silicon Monolithic Active Pixel Sensors (MAPS) covering a total area of approximately \(7\,\text{m}^2\). These sensors feature a rectangular pixel size of \(50 \times 150\text{ }\mu\text{m}^2\) and will have a time resolution of \(\sim 1\,\text{ns}\). The TDR is expected in Q4 2026.

Mighty tracker (MT) : To mitigate high occupancies and radiation levels, the MT has two distinct regions of complementary technologies, the inner region (Mighty Pixel) and the outer region (Mighty-SciFi). These detector regions are  
grouped into three separate tracking stations, T1, T2, and T3.  The Mighty Pixel is constructed of silicon pixel detectors featuring a total of six layers per station, covering a total area of approximately 8\,m$^2$ across 240 modules. The  pixels are fabricated in High-Voltage MAPS technology, each pixel of area  \(84 \times 84\text{ }\mu\text{m}^2\).  The outer region (Mighty-SciFi) consists of four planes of scintillating fibre detectors per station read out by Silicon Photomultipliers (SiPMs), giving 12 layers in total. The  four planes are each equipped with 250\,$\mu$m diameter fibres, oriented at angles of 0$^\circ$, -5$^\circ$, +5$^\circ$, and 0$^\circ$, respectively. To optimize performance, micro-lenses are integrated onto the SiPMs to enhance light collection by an expected 20\%. The SiPMs require a cooling system to maintain an operating temperature of approximately -100$^\circ$C.  
The TDR for the MT ensemble  is  expected in Q4 2026

{\bf RICH detectors}:
Hadron PID is achieved using two Ring Imaging Cherenkov detectors. RICH\,1, containing \(\text{C}_4\text{F}_{10}\) gas, covers the low-to-intermediate momentum region from \(\sim \)\,2 to 40\,GeV/c within an angular acceptance of 25 to 300\,mrad. RICH\,2 uses \(\text{CF}_{4}\) gas to extend the PID range from \(\sim \)\,15 to 100\,GeV/c, operating within an angular acceptance of 15 to 120\,mrad.
The upgraded system aims to improve or maintain the PID performance achieved in Run 3. To keep channel occupancy below 30\%, the system requires highly granular photodetectors that also feature a time resolution per single photon of 100\,ps or better, and SiPMs serve as the baseline technology. The primary advantages of SiPMs include an excellent photon detection efficiency (PDE), fine granularity (\(1\times1\text{ mm}^2\)), and good timing performance. However, SiPMs present challenges regarding their high dark-count rates, which necessitates cryogenic cooling, as well as yet  unproven radiation sensitivity when measuring single photons. RICH\,1 will undergo a re-build, utilizing a flat carbon-fibre mirror system  located inside the spectrometer acceptance to improve the chromatic aberration for imaging the Cherenkov rings.    The RICH-system TDR is expected in Q4 2026.

{\bf Electromagnetic Calorimeter (picoCAL)}:
The picoCAL design, optimized for $\pi^0$ and photon identification, must retain the  Run 3 electromagnetic energy resolution of \(\sigma (E)/E \approx 10\% / \sqrt{E} \,\oplus\, 1\%\), whilst providing a precise timing resolution of $\sim$20\,ps and  improved granularity. The high occupancy and radiation levels require a modular structure with multiple technologies across the different picoCAL regions.
The inner region will utilize a Spaghetti Calorimeter (SpaCal) technology featuring scintillating crystals and a tungsten absorber. The area exposed to 4 to 20\,Mrad of radiation will use 3\,$\times$\,3\,cm$^2$ cells, whilst the highest inner radiation zone experiencing up to 100\,Mrad will implement finer 1.5\,$\times$\,1.5\,cm$^2$ cells. The outer region will employ Shashlik technology, combining 2k existing modules with 1.3k newly-produced ones.
Regarding the Hadronic Calorimeter (HCAL) implemented up to Run 3, this would have minimal impact on the physics following the switch to a fully software-based trigger and therefore will be removed during Long Shutdown 3 (LS3) and not be replaced. The picoCAL TDR is expected to be ready by Q4 of 2026.

{\bf Muon stations}:
Again, the muon system will have an inner and outer region. The inner region  will be upgraded with 144 chambers of a novel Micro-Pattern Gaseous muRWELL detector technology \cite{Bencivenni_2020}, which can tolerate particle rates exceeding 1\,MHz\,cm$^{-2}$. For the outer region, the system will retain the present Multi-Wire Proportional Chambers (MWPCs), reusing 880 existing chambers with 80 newly-manufactured ones of higher granularity. The upgrade will also introduce new majority logic electronics  which will be very effective in maintaining the equivalent Run 3 performance of the system. The muon-system TDR is expected in Q4 2026.

{\bf DAQ and Trigger}:
LHC Run 5 will utilize a fully software-based trigger system through the implementation of Real Time Analysis (RTA), as for Run 3, but with data throughput increased by a factor of roughly 5 to 6. The system must maintain reasonable costs while ensuring robust data processing, comprehensive calibration capabilities, and fast alignment turnaround. Based on experience from Upgrade I, Graphics Processing Units (GPUs)  offer the most economical solution for the first-level of the High-Level Trigger (HLT1). Upgrade I relied on Central Processing Units (CPUs) for the second level  (HLT2), however  ongoing R\&D is  exploring the possibility to switch  to GPUs for \mbox{Upgrade II} for cost reduction. An  output rate to storage of approximately a  10\,GByte\,s$^{-1}$ is anticipated from HLT2 in Run 5.  For pragmatic reasons of expected computing evolution, the 
RTA and computing TDRs will be delayed with respect to the other TDRs by several years.

\section{LHCb Upgrade II planning}

The Upgrade II project has evolved from the Expression of Interest in 2017 to the Scoping Document in 2025. 
Currently  the project is in its TDR phase, and submission is expected for each of  the  subsystems mostly by the end of 2026. The construction phase
will proceed from 2027 to 2033.  In parallel to construction,  LHCb Run 4 will maintain the running luminosity at $\sim 2\,\times$\,10$^{33}$\,cm$^{-2}$s$^{-1}$ but with an enhanced RICH system incorporating timing capabilities. LS4 between 2034 to 2035 will see installation of the new detector and \mbox{Run 5} onwards from 2036 will  start the physics running and  exploitation.

The baseline detector is currently costed at 156\,MChF, with a reduced scope option at 125\,MCHF. Compared to the baseline, the latter scenario results in reduced acceptance and some degradation in detector performance, reduced signal-to-noise ratios, and hence some degraded physics performance. CERN’s Research Board has given its strong encouragement for the Upgrade II baseline scenario  \cite{CERN-DG-RB-2025-535} and, subject to resources, formal approval is anticipated later this year.   
Recent funding uncertainties have forced a shift toward a reduced-scope scenario, prompting several ongoing optimization studies. These evaluations focus on reducing acceptance in the high-\(\eta \) region, the consideration of a cheaper technology for VELO mechanics, increased photon-detector reuse in the RICH system, implementing single readout in most PicoCal modules, and removing magnet stations and TORCH subdetectors (not discussed in this paper). Initial findings indicate that the impact on physics performance can be minimised, preserving the unique and exciting core of the LHCb Upgrade II   programme.

\section{Summary}

The LHCb experiment has performed spectacularly well, recording \(9\text{ fb}^{-1}\) of data during Runs 1 and 2, and achieving smooth operation in Run 3 with a new Upgrade I detector that trebled the sample size to a total of \(37\text{ fb}^{-1}\). Looking forward, operating the experiment during the high-luminosity phase of the LHC will enable highly precise measurements in the flavour sector, extending the reach by a factor of up to 10 beyond the current accumulated data.
 LHCb Upgrade II  is an ambitious project with excellent prospects, driven by an intense R\&D programme that pushes the boundaries of existing particle-physics technologies. The project has received strong encouragement from the CERN Research Board, with detector optimisation and funding negotiations steadily progressing. Backed by an exciting physics programme, the collaboration continues to grow and warmly welcomes new collaborators.


\section*{Acknowledgements}

The author would like to thank the organisers of Beauty 2026 for making the conference in Maastricht so  enjoyable, and successful. In particular, the author would like to thank the international conference chairs Robert Fleischer and Guy Wilkinson, and the efficient and hospitable local team led by Marcel Merk and Keri Vos.






\addcontentsline{toc}{section}{References}
\bibliographystyle{LHCb/LHCb}
\bibliography{main}

\newpage

\end{document}